%% file: paper.tex
\documentclass[letterpaper,twocolumn,10pt]{article}
\usepackage{usenix2019_v3}

\usepackage{formatting/shortcuts}

\makeatletter
\def\@copyrightspace{\relax}
\makeatother

\author{{\rm Karan Grover$^{\dag}$\thanks{Lead authors ordered alphabetically.} , Shruti Tople$^{\dag}$\footnotemark[1] ,  Shweta Shinde$^{\ddag}$, Ranjita Bhagwan$^{\dag}$ ,  and   Ramchandran Ramjee$^{\dag}$} \\ \\ $^{\dag}$Microsoft Research \qquad $^{\ddag}$UC Berkeley}

\begin{document}
\title{\codename: Practical and Secure DNN Inference with Enclaves}

\maketitle

\input{paper-body}

\bibliographystyle{IEEEtran}
\bibliography{paper}

\end{document}

%% file: paper-body.tex
\input{chapters/abstract}
\input{chapters/intro2}
\input{chapters/problem}

\input{chapters/attack}
\input{chapters/leakage-source}

\input{chapters/design}
\input{chapters/evaluation}
\input{chapters/related}

\input{chapters/conclusion}

%% file: chapters/abstract.tex
\begin{abstract}

Cloud providers are extending support for trusted hardware primitives
such as Intel SGX. Simultaneously, the field of deep learning is
seeing enormous innovation as well as an increase in adoption.  In
this paper, we ask a timely question: ``{\em Can third-party cloud
services use Intel SGX enclaves to provide practical, yet secure DNN
Inference-as-a-service?}'' 
We first demonstrate that DNN models executing inside enclaves are
vulnerable to access pattern based attacks. We show that by simply
observing access patterns, an attacker can classify encrypted inputs
with {$\bf 97$\%} and {$\bf 71$\%}   attack accuracy for MNIST and
CIFAR10 datasets on models trained to achieve $99$\% and $79$\%
original accuracy respectively. This motivates the need for \codename,
a system we have designed for secure, 
easy-to-use, and performance efficient inference-as-a-service. 
%
%
\codename is {\em input-oblivious:} it transforms any deep learning
framework that is written in C/C++ to be free of input-dependent access
patterns thus eliminating the leakage.
\codename is {\em fully-automated and has a low TCB:} with zero
developer effort, given an ONNX description of a model, it generates compact and enclave-compatible code which can be deployed 
on an SGX cloud platform. 
\codename incurs {\em low performance overhead}: we use \codename with
Torch framework and show its overhead to be {$\bf 17.18\%$} on average
on $11$ different contemporary neural networks.

\end{abstract}

%% file: chapters/intro2.tex
\section{Introduction}
\label{sec:intro}

In recent years, deep neural networks (DNNs) have revolutionized
machine-learning tasks such as image classification, speech
recognition, and language
translation~\cite{lecun2015deep,hinton2012deep,long2015fully}.  Today,
the idea of applying DNNs to applications such as medical health
prediction and financial modeling hold tremendous promise. Consider a
medical health enterprise, Acme Corp., that has developed a
state-of-the-art DNN-based model for identifying diseases from
radiological images. Acme Corp. does not have its cloud
infrastructure but wants to monetize this model by making it
available to hospitals worldwide using a third-party cloud provider.
Acme wants to keep the model parameters (i.e., weights and biases)
confidential since it is precious intellectual property. On the
other hand, hospitals  want to keep their radiology data secure given
their privacy-sensitive nature.  However, both Acme and its clients
would still want to benefit from the cost-efficiencies of the cloud,
and the functions it enables. Thus, a {\it secure deep learning
inference service} is critical for addressing such scenarios.

Machine learning models hosted on the cloud are susceptible to several
types of security and privacy
issues~\cite{tramer2016stealing,fredrikson2015model,shokri2017membership,usenix16,mlcapsule,chiron,slalom}.
In this work, we focus specifically on preserving the {\em
confidentiality} of the model parameters and the  inputs to the model
from an attacker who can compromise the cloud software stack, or  can
gain physical access to the DRAM or the address bus. To protect code
and data from compromised cloud software and malicious administrators,
cloud providers are extending support for trusted hardware primitives
such as Intel SGX~\cite{asylo,msr-azure}. SGX supports
hardware-isolated execution environments called {\em enclaves} that
ensure that the user-code and data are inaccessible even to the cloud
administrators. Nevertheless, we are still far from realizing an
end-to-end DNN inference service using SGX enclaves due to the
following three important challenges:

\noindent
{\bf Security.} 
Prior work has shown that enclave memory is susceptible to leakage via
data access
patterns~\cite{pigeonhole,controlled,lee2016inferring,tople2017trade}.
However, it is not obvious that DNN model inference, that mostly uses
matrix multiplications,  is susceptible to such attacks. In
Section~\ref{sec:attack}, we motivate the need for data-oblivious DNN
implementations  within enclaves by demonstrating a {\it
first-of-its-kind attack}. We show that  an adversary can predict the
input label with reasonably high accuracy (close to the original model
accuracy) simply based on the observable access patterns during the
execution of the model. Thus, we require the inference service to be
{\em resistant to such memory address access pattern based  attacks}.

\noindent
{\bf Ease-of-use.} 
Developing custom applications for enclaves is cumbersome. This is
mainly because SGX does not provide system-call support, dynamic
threading, etc.~\cite{haven, scone, panoply}. Thus, the inference code
has to be carefully partitioned into inside-enclave and
outside-enclave code. Moreover, programming tools and debuggers that
run within SGX are rudimentary~\cite{sgx, sgx-explained}. The
alternative is to run entire applications that have been written for
an insecure setting, unmodified, on entire operating systems and
sandboxes that run within the enclave. Unfortunately, this results in
a large trusted computing base (TCB)~\cite{haven,graphene} which is
undesirable for security-sensitive applications. Ideally, the service
should require {\em no custom programming} and yet, it should have a
{\em low TCB}.

\noindent
{\bf Performance.} 
Finally, the solution can be practical only if the performance
overheads are ``acceptably'' low.

Unfortunately, previous research on secure DNN inference using SGX
does not solve these challenges entirely.  Prior work has either
concentrated on customized algorithms and code that can fix
access-based attacks at the cost of ease-of-use~\cite{usenix16}, or on
solutions that are easy to use but do not protect against access-based
attacks and have a large TCB~\cite{chiron}.  To make matters worse,
these approaches have not been extensively evaluated on contemporary
DNN models to ensure that performance overheads are indeed uniformly
low. To this end, we present \codename, a secure DNN inference service
that simultaneously provides obliviousness, low TCB, ease-of-use, and
low-performance overhead.\footnote{\codename is a Spanish word for
{\em a confidential friend or a confidant}.} \codename consists of two
main components---\codename-Converter and \codename-Generator,  that
help us provide these properties. To our knowledge, \codename is the
first system that targets and solves all these challenges in a single
system.

We first address the {\em security} challenge.  The key observation
that enables \codename's approach  to data-obliviousness is that deep
learning models predominantly have data-independent accesses and very
few data-dependent accesses (Section~\ref{sec:insight}). Our main
contribution is the observation that deep learning models exhibit very
specific and regular data access patterns that we call {\em \pattern}
patterns. In programs with these patterns, every input-dependent
conditional branch either assigns a value to a variable otherwise
exits the branch. The presence of such a specific pattern allows us to
completely automate the task of eliminating these memory accesses
using simple yet efficient oblivious primitives, thereby making the
inference process resistant to access-based attacks. Prior works have
proposed manual modifications to neural network algorithms, thereby
restricting the applicability of the solution to a handful of
models~\cite{usenix16}.\footnote{Similar to the threat model in
previous work~\cite{chiron, mlcapsule, slalom}, we do not provide
protect the model structure such as the depth, neurons in each layer,
etc.} 

We present \codename-Converter---a tool that automatically detects all
such data-dependent access patterns in a given deep learning
framework. It modifies them to become data-independent using any of
the known oblivious constructs~\cite{raccoon,zerotrace,usenix16}. In
our implementation, we select the $\tt{CMOV}$ instruction for this
purpose. With \codename-Converter, we  transform the popular
Torch framework to guarantee obliviousness~\cite{Torch}. This makes
\codename expressive to a large number of state-of-the-art models, as
shown in our evaluation in Section~\ref{eval}. However, our
observations about the access patterns in neural networks are
independent of the framework and are applicable to deep learning
algorithms in general. 

To address the {\em ease-of-use} challenge, \codename uses the
\codename-Generator which takes models represented in the popular Open Neural Network eXchange (ONNX)
format~\cite{onnx} as input and automatically generates a minimal set
of enclave-specific code and encrypted parameters for the model. 
\codename does not require any custom programming and thus has zero
developer effort. The server simply loads the auto-generated code for
the model within an enclave where the model parameters are decrypted.
To reduce the TCB size, the \codename-generator only includes files
required for building the model instead of the entire DL framework. 
This approach, we find, reduces the TCB size by {$ \bf 34.7\%$}.
Finally, we address the {\em performance} challenge. We evaluate 
\codename on {$\bf 11$} models used in real-world deep learning cloud
services and show that \codename has {$\bf 17.18\%$} overhead on
average for inference for MNIST, CIFAR10, and ImageNet image datasets.

In this paper, we make the following contributions:
\begin{itemize}
\squish
\item {\em Access Pattern Based Attack.}	
We demonstrate a concrete attack that  predicts the output class of
encrypted inputs to the neural network based on the access patterns
observable in the intermediate layers. Our attack achieves {$\bf
97$\%} and {$\bf 71$\%} accuracy for predicting MNIST and CIFAR10
images on an MLP and LeNet model  respectively.

\item {\em End-to-end Design.} 
Starting with an ONNX model, we show that our system auto-generates
data-oblivious code which  runs seamlessly on SGX in the cloud with
zero development effort for the cloud customer.

\item {\em \codename System.} 
We present \codename, which incorporates the above-mentioned
constructs into the Torch framework.  \codename converts Torch to
replace data-dependent branches and it auto-generates minimal amount
of Torch code from the ONNX model specification, thereby ensuring that
the TCB size is small. Our evaluation shows that \codename incurs an
average performance overhead of {$\bf 17.18\%$} on $11$ models.

\end{itemize}

%% file: chapters/problem.tex
\section{Problem}
\label{sec:problem}

In this section, we first outline the problem setting for \codename.
We then describe our threat model. Finally, we outline the desirable
properties that \codename should achieve.

\subsection{Secure Inference-as-a-service}
\label{sec:setting}

Neural network algorithms operate in two phases---training and
inference. The training step uses a labelled dataset to generate {\em
model parameters} i.e., weights and biases such that the error between
the true input class and the predicted class is minimum. Once a
network is trained, the inference phase uses the model parameters to
accurately predict the output class for any given input. We assume
that the training happens a priori in a trusted environment while the
inference is offered as a cloud service to benefit other users or for
monetary gains. To ensure confidentiality guarantees, such a
cloud-based inference service should (a) protect the model parameters
from the server and the users, and; (b) protect the users' inputs and
outputs from the server and other users of the service.

\input{figures/setting.tex}

Figure~\ref{fig:setting} shows the entities involved in such an
inference service: the {\em cloud provider}, the {\em model owner},
and multiple {\em model users}.  We use the term model users and users
interchangeably.  The cloud provider supports trusted hardware
primitives such as Intel SGX. SGX-enabled CPUs create isolated
execution environments called {\em enclaves}  {\em where all the code
and data is protected from direct access via the untrusted software
such as the  OS as well as physical adversaries.} SGX allows remote
attestation of the code running inside the enclaves to ensure its
integrity~\cite{sgx2}.  

\paragraph{Uploading \& Instantiating the Model.}
A model owner has two components: (a) a model binary that contains 
the code corresponding to the model architecture; and (b) the model
parameters (i.e., the weights and biases) generated after training the
model architecture on a sensitive training dataset. The model binary /
model architecture is public i.e., it is known to other entities such
as the cloud provider or the  users. On the other hand, the model
parameters are the key intellectual property and hence are encrypted
before sending to the server.

The model owner first uploads the model binary and attests the
correctness of the loaded binary inside the enclave. After a
successful remote attestation, the model owner establishes a secure
channel with the enclave that terminates inside the
enclave~\cite{ra-tls}. Concretely, the model owner and the enclave
perform a standard Diffie-Hellman key exchange to establish a shared
secret key. Using this secure channel, the model owner provisions the
model parameters to the enclave. When instantiating the model, the
enclave decrypts these parameters using the shared key. This completes
the process of hosting the inference service on the cloud.

\paragraph{Querying the Service.}
After the trusted inference enclave is set up, each user of this
cloud inference service remotely attests the enclave to verify the
correctness of the code executing inside it via standard methods as in
other  Intel SGX based cloud solutions~\cite{haven, ryoan}.  Once the
attestation is complete, the user establishes a separate secure
channel with the enclave to provision secrets.  During this step, each
user generates a distinct shared secret key with the enclave.  To
query the  inference service, users encrypt their inputs with their
respective secret keys and send them to the enclave. The enclave
decrypts the input, runs the model owner's binary on the user' input,
and  encrypts the prediction output with the secret key corresponding
to each user. Users receive the encrypted output label that can be
decrypted only with their secret key.  Throughout the process, the
untrusted cloud software learns nothing about the model parameters
(property of the model owner) or the input/predicted output (property
of the model user). Further, the  model owner learns nothing about the
model users input/output and the model user learns nothing about the
model parameters. As each user shares a separate key with the enclave,
it is guaranteed that each user's input remains confidential from all
the other users.

\subsection{Threat Model}
\label{sec:tm}

In our threat model, we consider the cloud provider to be untrusted
since an  adversary can exploit bugs in the cloud software stack and
get privileged access to the entire system. The only trusted entity is
the SGX-enabled CPU processor available on the server. SGX guarantees
that the hardware-isolated enclaves are inaccessible to the adversary.
Enclaved execution guarantees that the encrypted content within the
enclave gets decrypted only inside the processor package. 

Although SGX prevents direct access to the enclave memory region, an
adversary can learn significant information about the sensitive data
via side-channels, specifically the access patterns. We consider
software as well as a physical adversary who access the DRAM or  can
snoop over the address bus to learn the memory accesses and get the
execution trace of the enclaved application. Also, the adversary can
perform page fault and cache-channel attacks to learn the execution
flow within the
enclaves~\cite{pigeonhole,controlled,shih2017t,van2017telling,gras2018translation,oleksenko2018varys,brasser2017software,gotzfried2017cache,chen2017detecting,fu2017s}.
Moreover, the adversary can monitor the system call interface and
observe access  to the storage disks and trace network
packets~\cite{tople2017trade}. 

\paragraph{Assumptions.}
Similar to prior work in this
domain~\cite{usenix16,mlcapsule,chiron,slalom}, we assume the
adversary knows the hyper-parameters of the model such as the type of
training data (e.g., images, text), model architecture, the number of
layers, the number of neurons in each layer; these do not leak
information about the sensitive inputs. We aim to provide secure
inference service on a model without revealing the model parameters or
the  query input. Although leakage is possible via another
side-channels such as timing, we do not address them in this work. 
However known solutions to mitigate timing channels can be used with
\codename~\cite{coppens2009practical,cleemput2012compiler,barthe2006preventing}.
Other types of attacks such as denial-of-service are not within the
scope of this work. We assume that the cloud provider would always
respond to the model users to maintain its reputation. Lastly, we
assume that all the SGX guarantees are preserved i.e., there is no
hardware backdoor present in the processor package and the secret keys
are not compromised via any attack. We assume that remote attestation
step and establishing of the secure channel are secure and are
performed as per standard guidelines~\cite{sgx3}.


\paragraph{Out-of-scope: Privacy Attacks.}
A plethora of privacy-related attacks on machine learning models have
been demonstrated  such as model-inversion~\cite{fredrikson2015model},
model stealing~\cite{tramer2016stealing}, membership
inference~\cite{shokri2017membership} and others. These attacks  focus
on compromising the privacy of either the training dataset or the
model parameters. Model inversion and membership inference attacks
target the training dataset while model stealing attacks aim to
approximate the model parameters by observing the prediction score of
the output. Although leakage via these attacks is an important
concern, designing solutions to mitigate them is orthogonal to the
main goal of \codename.  In \codename, our primary focus is to ensure 
{\em confidentiality} of the original model parameters and the
encrypted inputs of users while enabling a multi-user cloud-based
inference service.  \codename does not ensure protecting  the privacy
of the data-points involved during the training step and hence is
susceptible to these attacks. However, a model owner can use existing
solutions such as  differentially-private learning techniques or
adversarial learning during the training phase before uploading the 
model to \codename
system~\cite{abadi2016deep,papernot2016semi,nasr2018machine}.

\subsection{Desirable Properties}
\label{sec:properties}

\codename aims to satisfy the following desirable design properties
that are necessary to create a practical and secure inference service.

\begin{itemize}
\squish
\item  {\em Input-Obliviousness:} 
We aim to achieve oblivious access patterns during the execution of the
inference phase, thus preventing data leakage due to observable
data-dependent memory address accesses. 

\item {\em Low-TCB:} 
The trusted computing base for any model should be the smallest
possible subset of the entire deep learning framework to reduce the
attack surface from bugs or vulnerabilities in the code.

\item {\em Zero Developer Effort:} 
With \codename, users need not write custom SGX-specific code. Any
machine learning scientist should be able to use \codename out of the
box.

\item {\em Expressiveness:} 
We aim to support inference on state-of-the-art neural network models
that contain complex combinations of linear and non-linear layers with
parameters that range up to tens of millions.

\item {\em Backward Compatibility:} 
We aim to support inference for models that have been trained in the
past using any of the existing deep learning frameworks (e.g., Caffe,
TensorFlow, PyTorch, CNTK).

\item   {\em Performance:} 
Lastly, we expect \codename to have low performance overhead across
several models trained  on different input datasets.

\end{itemize}

%% file: figures/setting.tex
\begin{figure}[t]
\centering
\includegraphics[scale=0.8]{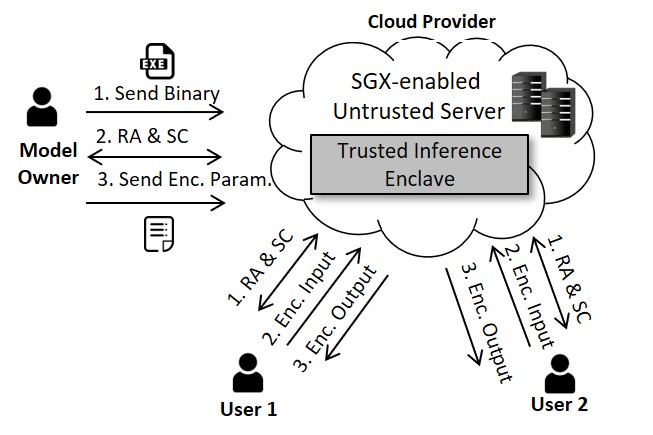}
\caption{ 
A secure inference-as-a-service setting. The model owner (1) sends the
model binary, (2) performs remote attestation (RA) and establishes a
secure channel (SC), and  (3) sends the encrypted model parameters to the enclave. The model users (1) perform RA \& SC, (2) send the encrypted inputs for inference, and (3) receive encrypted outputs.
}
\label{fig:setting}
\end{figure}

%% file: chapters/attack.tex
\section{Access Pattern Based Attack}
\label{sec:attack}

We demonstrate that an adversary can indeed predict the labels of
encrypted input data by merely observing access patterns of a DNN. 
Our high attack accuracies motivate the need to eliminate the leakage
via memory access in the secure cloud-based DNN inference. 

\subsection{Target Model and Insight}
\label{sec:attack_ovw}
We describe the attack strategy and perform the  attack on two
network models trained on MNIST and CIFAR10 datasets respectively. Our
attack predicts the labels for encrypted inputs with {$\bf 97$\%} and
{$\bf 71$\%} accuracy, respectively.

\input{figures/attack_steps}

\paragraph{Target Model Architecture.}
We demonstrate our attack on two target deep neural network models
trained using the Torch framework~\cite{Torch}.  The first model is a
multi-layered perceptron (MLP) model with $4$ layers trained on the
MNIST dataset~\cite{mnist}.  MNIST is a collection of $32
\times32$-pixel black and white images of hand-written digits
($0$-$9$).\footnote{We used the Torch framework that provides
$32\times 32$-pixel images instead of the common $28 \times 28$ for
MNIST dataset.} We have trained the model to achieve $99$\% accuracy.
Next, we train a LeNet model which is a convolutional neural network
(CNN) with $3$ convolution layer followed by $2$ fully-connected
layers. The model is trained to achieve $79$\% prediction accuracy on
CIFAR10 dataset~\cite{cifar}. The CIFAR10 dataset has a collection of
colored images for 10 different classes. Each hidden layer in both the
models is followed by a rectified linear activation function (ReLU)
layer. ReLU is a commonly used activation function to train accurate
models. We perform our attack based on memory accesses observed only
at these ReLU layers. Lastly, note that the final layer of both these
models is a linear layer followed by a softmax function that outputs
the probabilities of various classes. Note that we select MNIST and
CIFAR10 as our example benchmarks, however the attack is feasible on
any other dataset (such as ImageNet) trained on any model with
data-dependent layers in the network.

\paragraph{Leakage-prone Layer.}
The ReLU function is represented as $\tt max(0,x)$ which is commonly
implemented using a conditional branch in several ML frameworks such
as
PyTorch\footnote{https://github.com/torch/nn/blob/master/lib/THNN/generic/Threshold.c},
Caffe\footnote{https://github.com/BVLC/caffe/blob/master/src/caffe/layers/relu\_layer.cpp},
and
Tensorflow\footnote{https://github.com/tensorflow/tensorflow/blob/master/tensorflow/core/\\kernels/relu\_op\_functor.h
}. Listing~\ref{lst:relu} shows the code for the implementation of
ReLU layer in the Torch library. The code is present in the
$\tt{Threshold.c}$ file which provides a general threshold
functionality  and is not limited to the value zero. 
\input{listings/threshold} 
The function simply activates the neurons with values greater than
zero (or threshold) and deactivates others.  Consequently, the memory
access pattern differs for every input based on whether the branch is
executed or not. Therefore, observing the access patterns at  the ReLU
layer, an attacker can distinguish between neurons that are activated,
and neurons that are not.  In our attack, we exploit the relation
between the activated neurons and the class label for any given input,
without explicitly learning the exact neuron activations.

\input{chapters/attack_details}

\subsection{Attack Results} 
\label{sec:attack_results}

The attacker's classifier achieves a $9$-fold cross-validation accuracy
of $97$\% and $71$\%  for MNIST and CIFAR10 dataset respectively when
trained on a dataset size of $10,000$ inputs. This is the best-case
accuracy when the attacker can query maximum number of inputs to the
target model. We observe that the attacker accuracy is reasonably
close to the original model's accuracy i.e., $99$\% and $79$\% for
MNIST and CIFAR10, respectively. Figure~\ref{fig:mnist_attack} and
~\ref{fig:lenet_attack} gives the attack accuracy for predicting the
labels of test dataset of MNIST and CIFAR10 images when the attacker
classifier is trained over a range of training inputs. 

Note that the training features correspond to the memory access
patterns of the last layer and the last two ReLU layers. This shows
that the access patterns observed at the ReLU layer are critical and
have the potential to reveal information about the sensitive inputs.
These access patterns essentially capture the neuron activation
information in these layers which is directly related to the output
class.  Thus, all the inputs that belong to the same class have
similar address access pattern in the last layers. Note that, the
attacker does not need to know the mapping of the neuron activation to
the address patterns. The attacker classifier simply learns the memory
address access pattern corresponding to each output class. This result
can be explained from the understanding in machine learning theory
that the last layer is the most representative of the input class
(since this is what is used by the softmax layer to output class
probabilities)~\cite{shokri2017membership}. Thus, an attacker who
observes the access patterns only at the last layer is still capable
of guessing the labels for the encrypted inputs with good accuracy. 

\input{figures/attack_results}

We train the attacker classifier on access patterns with different
number of query inputs. We observe that the attack accuracy is high
for MNIST data even for smaller samples while it gradually increases
for CIFAR10 with increasing in the training data size. This is due to
the difference in the complexity of features in these two datasets.
Thus, we observe that attacking a model trained on complex features
would require larger number of queries to generate the inputs features
for the attacker classifier. Consequently, when an honest user sends
an encrypted image to the cloud, the adversary can apply  the
classifier on the observed access patterns and  determine the input
label of the encrypted image with high confidence. Therefore, it is
critical that access patterns for DNN inference be made oblivious to
prevent these attacks. 

To conclude, we demonstrate very high attack accuracy in a worst-case
attack scenario where the attacker has access to the entire memory
address trace. However, similar attacks can be performed using either
page-fault~\cite{pigeonhole,controlled,van2017telling} or cache
side-channel~\cite{gras2018translation,brasser2017software,gotzfried2017cache}
to collect features corresponding to the branching of data-dependent
conditional statement.

%% file: figures/attack_steps.tex
\begin{figure*}[t]
\centering
\includegraphics[scale=0.6]{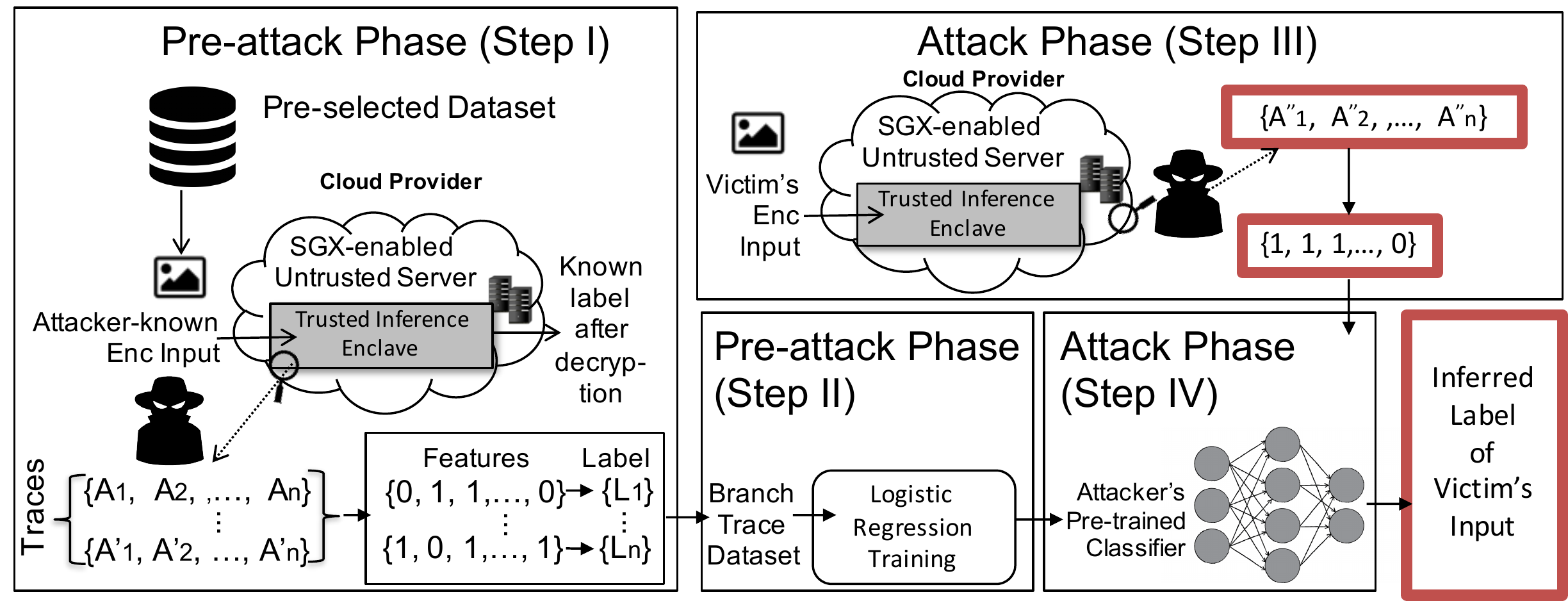}
\caption{Attack steps where attacker:  
(I) executes its own dataset with known labels, collects  traces, and
extracts features; 
(II) uses the branch trace dataset (features, labels) for training a
simple model;
(III) observes the trace at runtime when the victim is executing 
inference on its own image inside the enclave, processes the trace to
extract the features; 
(IV) passes the features to the pre-trained model to guess the label
of victim's input. 
}
\label{fig:attack_steps}
\end{figure*}

%% file: listings/threshold.tex
\begin{lstlisting}[style=JavaScript, language=C, xleftmargin=0.3cm, 
caption={Input-dependent code  from $\tt{Threshold.c}$ in Torch.
},captionpos=b, label={lst:relu}]
      if (*input_data <= threshold)
        *input_data = val;
    );
\end{lstlisting}

%% file: chapters/attack_details.tex
\subsection{Attack Details}
\label{sec:attack_details}

Figure~\ref{fig:attack_steps} describes the key steps of our attack.
In summary, the attacker performs the following steps:
\begin{itemize}
\squish
\item {\bf (Step I).}
In the pre-attack phase, first, the adversary uses its own input
dataset to collect the memory access patterns and their corresponding
output class predicted by the trained target model.

\item {\bf (Step II).} 
The attacker feeds these memory patterns as input features to
train a classifier with logistic regression. 

\item {\bf (Step III).} 
During the attack phase, to learn the output labels of other users
(the victim), the adversary  collects the memory address trace during
the inference execution of the victim's input. 

\item {\bf (Step IV).} 
Finally, the attacker uses its trained classifier to predict the
output class of victim's input using the observed memory accesses.

\end{itemize}

\input{figures/attack_feature_trace}

\paragraph{Observing Memory Access Patterns.}
Here, we describe one concrete example of a memory access pattern
trace that the attacker (the untrusted OS in this case) can collect
via the page-fault channel. Since the OS is in charge of allocating
physical memory to the enclave, it can severely limit the number of
pages the enclave gets (say $1$ page)~\cite{pigeonhole}. Every time
the enclave accesses an address on a different page, it incurs a
page-fault which is delivered directly to the OS. The page-fault
information consists of the page number and not the offset within the
page (last three bytes of the address).
Figure~\ref{fig:attack_feature_trace} (b) shows the actual addresses
being accessed inside the enclave and
Figure~\ref{fig:attack_feature_trace} (c) shows the page faults
visible to the OS when executing LeNet model inference over two
different inputs with labels $\tt{airplane}$ and $\tt{horse}$
respectively. If the attacker uses some other channel or their
combination, in the worst case, it can learn the entire address (page
number as well as the offset within that page). We refer to any  of
these as the memory address trace for simplicity.

\paragraph{Extracting Input Features for the Classifier.}  
To extract the input features from the observed memory address traces,
the attacker can do a $\tt{diff}$ between multiple traces.  In our
example, Figure~\ref{fig:attack_feature_trace}, we see that most of
the memory accesses during the execution are deterministic. The
location where the trace begins to differ, highlighted by bold and
blue text in Figure~\ref{fig:attack_feature_trace} (b), indicates the
execution of data-dependent branch condition in the code.
Specifically, the address $\tt{0x8915de}$ (highlighted in bold)
belongs to the conditional statement that appears in both the traces.
The address accessed after this location depends on whether the
condition is true or false. More importantly, access to the address
$\tt{0xb7939c}$ causes a page-fault for page $\tt{0xb79}$ and reveals
that the branch condition is not satisfied because of the $\tt{if}$
path is not executed. Whereas, when the enclave does not page-fault 
for this address, it reveals that the branch condition is satisfied.

Thus, by observing the difference in the traces of inputs from
separate classes, the attacker not only learns the address locations 
for the data-dependent branch conditions in the code  (i.e., ReLU in
this model) but also whether the branch condition was satisfied or
not. The attacker then extracts the features from the access pattern
of each input. Specifically, for a given trace, it uses $1$ to
represent that the if branch executed and a $0$ otherwise. The
attacker does this for the entire trace which has multiple occurrences
of ReLU because of a loop around the branch condition. At the end, the
attacker has a vector of binary input features for each input image. 

\paragraph{Creating Labelled Training Dataset.} 
Our target model i.e., MLP or LeNet runs inside an enclave. The
attacker acts as a legitimate user and establishes a secure channel
with the enclave. Next, the attacker encrypts its input and queries
the target model for inference. During the enclave execution, the
attacker operating at the cloud provider records the memory address
trace for the requested input via one of the well-known Intel SGX
cache
side-channels~\cite{gras2018translation,brasser2017software,gotzfried2017cache}
or controlled-channel
attacks~\cite{pigeonhole,controlled,van2017telling} (see next
paragraph for details).  The enclave responds with an encrypted output
label to the attacker, who then uses its own key to decrypt the output
label. The attacker repeats this process for  various input images to 
collect memory address traces and corresponding labels. The attacker
uses the above feature extraction method to generate binary input
features corresponding to the memory access patterns. This becomes the
labelled training dataset for the classifier.

\paragraph{Training Attacker's Classifier.} 
We use a simple logistic regression-based model for training the
attacker's classifier, where the number of input features for each
input is equal to the total number of neurons in all the ReLU layers.
Using input features proportional to the number of neurons may
increase the computation cost for training the attacker classifier and
grow significantly large for bigger models with millions of
parameters. To reduce the number of input features in our attack, we
use only the features that correspond to the last layer of the model.
Since the hyper-parameters of the model are public, one can
selectively generate only last $n$ features from the memory address
trace.\footnote{We assume sequential execution trace of the model.}
Here, $n$ corresponds to the number of neurons in the last layer. To
understand if the attack accuracy improves with an increase in the
number of features, we trained another classifier with input features
generated from the last two ReLU layers of each of the two models. 
Further, we trained several instances of the attacker classifier for
both the MNIST and CIFAR10 dataset by varying the number of training
inputs. This allowed us to understand the maximum number of inputs
that the attacker needs to query the target model for building a
highly accurate classifier. This is useful to estimate the cost of the
attack as the inference service might charge the users for every
query. 

%% file: figures/attack_feature_trace.tex
\begin{figure*}[t]
\centering
\includegraphics[scale=0.55]{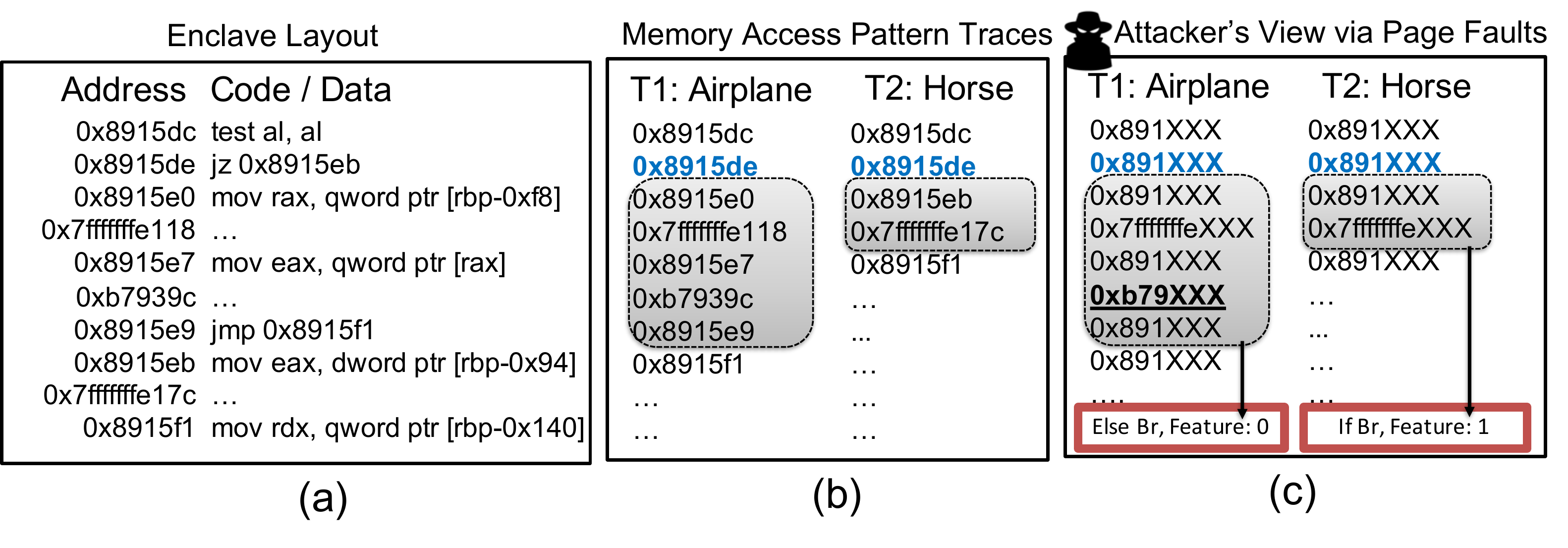}
\caption{
Enclave memory address layout and attacker's observations. 
(a) The enclave address and code layout for ReLU when executing LeNet
model.
(b) Two different address traces exhibited when the enclave is
executing LeNet model for inputs with labels airplane and horse
respectively. The highlighted part in the dotted gray box shows the
input-dependent difference in the trace. The bold blue text shows the
branching instruction.
(c) Page faults  observed by the attacker when it allocates only one
physical page to the enclave. When the attacker sees a page-fault for
the underlined and bold page number it marks the feature value
(indicated by box with thick border) to be $0$ ($\tt{if}$-branch not
taken); otherwise it marks the value to be $1$ ($\tt{if}$-branch
taken).
}
\label{fig:attack_feature_trace}
\end{figure*}

%% file: figures/attack_results.tex
\begin{figure}[t]
\centering
\begin{minipage}[h]{0.4\textwidth}
\epsfig{file=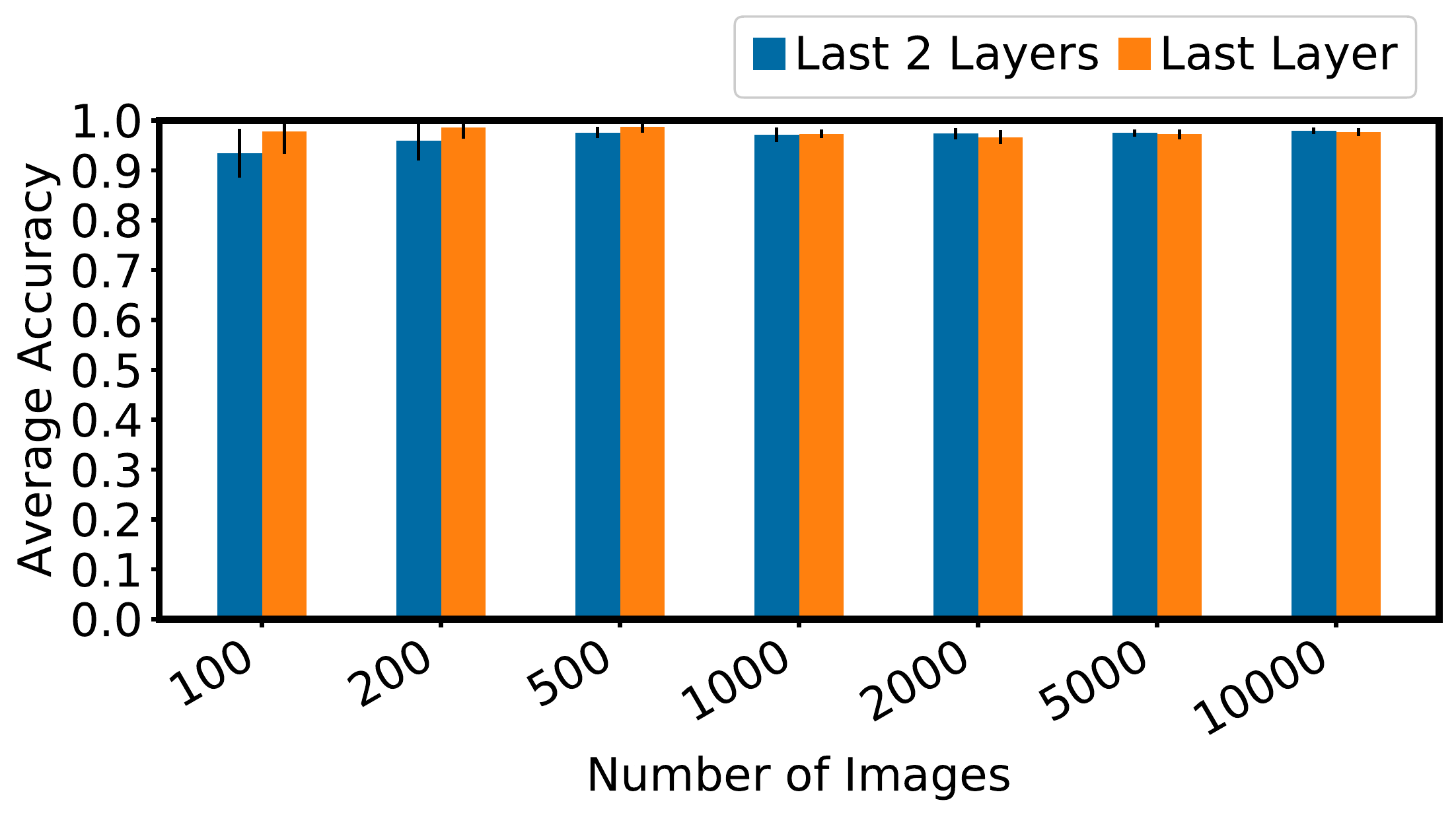, scale=0.28}
\subcaption{MNIST with a MLP model.}	
\label{fig:mnist_attack}
\end{minipage}
\begin{minipage}[h]{0.4\textwidth}
\epsfig{file=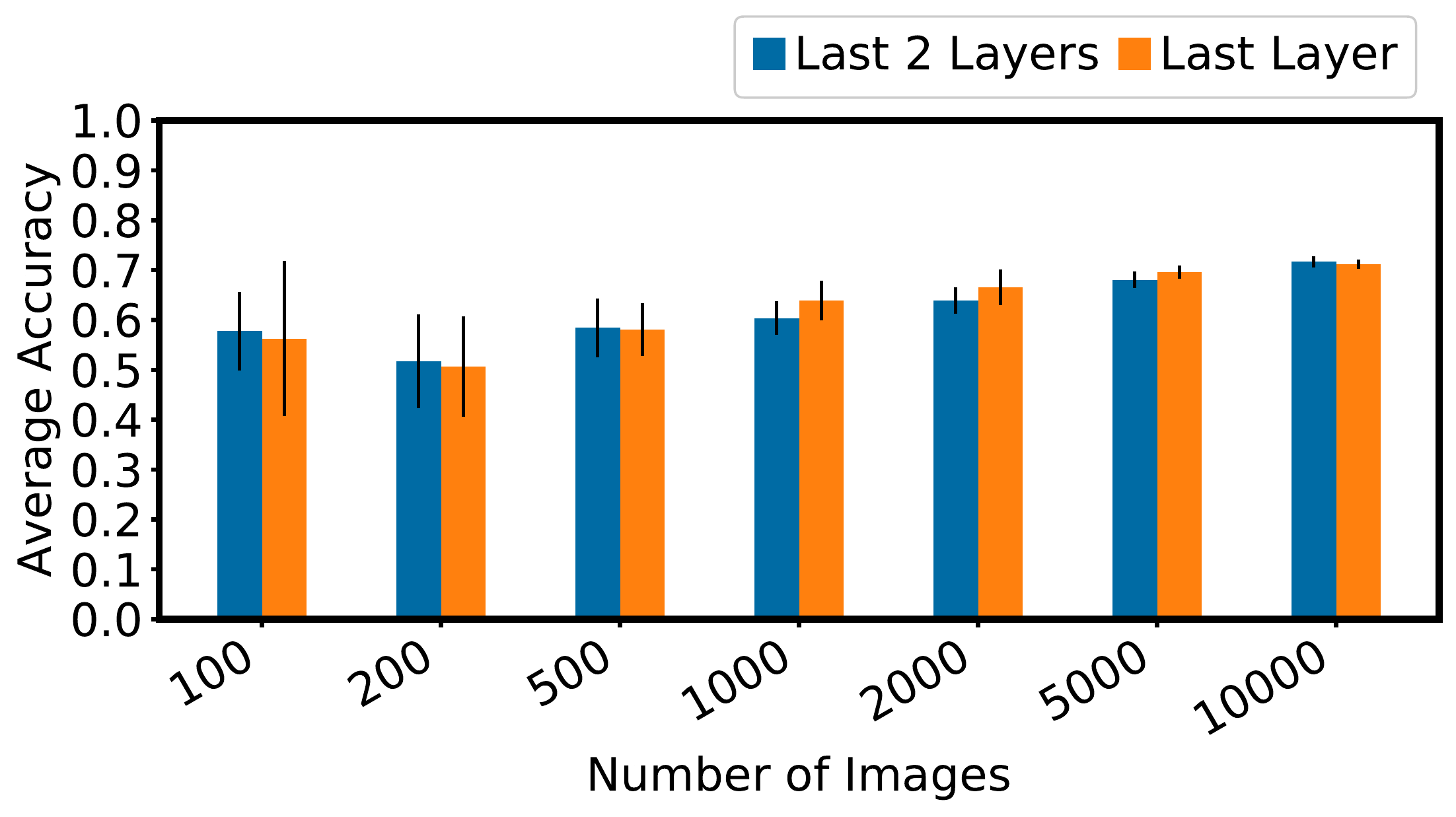, scale=0.28}
\subcaption{CIFAR10 with LeNet model.}	
\label{fig:lenet_attack}
\end{minipage}
\caption{Accuracy results for access pattern based attack performed using a logistic regression model.}
\label{fig:attack_results}
\vspace{-0.3cm}
\end{figure}



%% file: chapters/leakage-source.tex
\section{Sources of Leakage in DNNs}
\label{sec:insight}

Leakage via access patterns is an important concern when using SGX for
designing secure systems. Recently, researchers have proposed
Oblivious RAM based solutions for SGX to eliminate leakage for
arbitrary read-write patterns~\cite{zerotrace,obliviate}. However,
these solutions incur significant performance overhead that makes them
inefficient for a DNN inference service.

By customizing our solution to the specifics of how memory is accessed
in DNN algorithms, we believe the overhead of making DNN inference
data-oblivious can be significantly reduced. We make two key
observations. First, the vast majority of computations in DNNs involve
linear layers (fully connected or convolution layers) that exhibit
only {\em deterministic} accesses, i.e., the memory access patterns do
not vary based on the input. Second, certain types of DNN layers, such
as ReLU or max-pool, exhibit input-dependent accesses, i.e., their
memory access patterns can vary depending on the input (as we show in
Section~\ref{sec:attack}). However, even in layers that exhibit
input-dependent accesses, the accesses are of a very specific type: in
each branch either a given memory location is accessed or no memory
locations are accessed. We call these {\em \pattern} patterns. This
important observation with respect to the general class of deep
learning algorithms allows us to build an efficient system that is
resilient against access pattern based attacks. Further, this
observation mainly allows us to  automate the steps in \codename
without developer effort, thus distinguishing it from prior
work~\cite{usenix16,mlcapsule,chiron,slalom}. We discuss these
observations in detail for layers that are commonly used in popular
neural networks and provide supporting code having \pattern~ patterns
in the Torch framework. 

\subsection{Linear and Batch Normalization Layers}

In neural networks, such as multilayer perceptrons (MLPs) and
convolutional networks (CNNs), the dominant part of the computation
can be represented as {\it matrix-multiplication} between the weights
and the inputs to each layer.\footnote{Sometimes convolutions are
computed using fast-fourier transforms but they are also
input-independent.} In fact, over $90$\% of computation in many modern
networks are attributed to the convolution
operation~\cite{computation90}. Similarly, recurrent networks (RNNs,
LSTMs) are also dominated by matrix
multiplications~\cite{rnncomputation}. Matrix multiplication involves
performing the same operation, irrespective of input values. This
makes their access patterns input-independent.

Batch Normalization~\cite{batchnorm} is another popular layer used in
modern DNNs. Batch normalization, at inference time, adjusts the input
value by the expected mean and variance of the population (computed
during training). Thus, batch normalization does not perform any input
dependent access.

\subsection{Activation Layers}

Activation layers are important in neural networks for capturing the
non-linear relationship between the input and the output values.
Several activation functions are used in these networks to improve the
accuracy of the models. Sigmoid and tanh are the most basic activation
functions which are computed as $\frac{1 }{(1 + e^-x)}$ and $\frac{2}
{(1+e^-2x)} - 1$ respectively. As seen from the formulas, neither
function incurs any input-dependent access patterns. While RNNs still
use tanh activations, the standard choice in  recent MLPs and CNNs
have been the ReLU activation~\cite{relu}. ReLU defined as $\tt
max(0,x)$ uses an input-dependent conditional branch as we described
in Section~\ref{sec:attack} and hence leaks information from its
access patterns. {The Torch framework offers $18$ different activation
functions out of which $15$ exhibit \pattern~ patterns.  They include
$\tt{ELU}$, $\tt{Hardshrink}$, $\tt{Hardtanh}$, $\tt{LeakyReLU}$,
$\tt{LogSigmoid}$, $\tt{PReLU}$, $\tt{ReLU}$, $\tt{ReLU6}$,
$\tt{RReLU}$, $\tt{SELU}$, $\tt{CELU}$, $\tt{Softplus}$,
$\tt{Softshrink}$, and $\tt{Threshold}$.  

We consider HardTanh as an example. The functional definition of
HardTanh is: $f(x) = 1$, if $x > 1$,  $f(x) = -1$, if $x < -1$,  $f(x)
= x$, otherwise. Listing~\ref{lst:hardtanh} shows the source code
from the $\tt{HardTanh.c}$ file in the Torch framework.  For each of
the conditions, the branch only executes an assignment operation if
true, otherwise, there is no assignment. Therefore, $\tt{HardTanh}$ 
exhibits the \pattern~ pattern.  Further, observe that the condition
is sequentially executed for all the inputs (or indices). The
$\tt{if}$ condition is executed inside a $\tt{for}$ loop that iterates
over all the inputs.  All the loop bounds are public values such as
the model hyper-parameters which do not leak information about the
sensitive parameters and the user inputs.  Observe that there are no
nested branches that are conditioned on sensitive inputs (such as
weights and bias). The $\tt{inplace}$ variable is public and  its
value can be decided a priori. We do not encounter accesses to memory
locations with sensitive index.   All the other activation functions
exhibit a similar use of \pattern~ patterns in their implementation. 
\input{listings/hardtanh}

\subsection{Pooling Layers}
It is common to use pooling layers after convolution layers to reduce
the output size at each layer. Two popular variants of pooling layer
are max-pool and mean-pool. Typically, a filter of size $2 \times 2$
with a stride of $2$ is applied to the input of this layer. The
pooling function replaces each of the $2 \times 2$ region in the input
either with a mean or max of those values, thus reducing the output
size by $75$\%.

As the mean-pool function simply computes an average of the values
from the previous layer, it exhibits a deterministic access pattern
irrespective of the actual input to the model.  However, the max-pool
variant selects the maximum value in this $2 \times 2$ window and sets
it as the output. Listing~\ref{lst:example} shows the code that
implements a max-pool operation. 
\input{listings/example}
The input-dependent branch statement
writes to the sensitive input location only if the condition is true.
The loop terminating conditions are public values ($2 \times 2$) that
does not leak information about the input.  Hence, similar to the ReLU
function, the max-pool operation exhibits an \pattern~ pattern. The
Torch framework offers $3$ types of pooling functions
$\tt{SpatialMaxPooling}$, $\tt{TemporalMaxPooling}$, and
$\tt{VolumetricMaxPooling}$. All of them exhibit \pattern~ patterns.

\subsection{Softmax Layer}
Most of the networks conclude with a softmax layer that calculates the
probability for each of the output classes.  The softmax layer
performs a deterministic computation of calculating a value
corresponding to each output class. The class that has the highest
value is then returned as the predicted class. Although the softmax
layer is oblivious in itself, computing the maximum value among all
the classes requires an input-dependent branch condition to find the
max value, similar to the max-pool function.  For this, a sequential
comparison is performed over all the output values to find the class
with the maximum probability. Again, this falls into the category of
\pattern~ patterns.

\subsection{Summary}
The above analysis of all the common layers used in neural network
algorithms during inference  show that all the sensitive
input-dependent access patterns can be categorized into the \pattern~
pattern. Moreover, the dominant computation cost ($90+$\%) is
matrix-multiplications that are input independent. Thus, a custom
data-oblivious solution for DNNs that addresses only the input
dependent access pattern is likely to have low overheads. We do not
consider the training of a model and therefore avoid operations that
leak information during back-propagation.

%% file: listings/hardtanh.tex
\begin{lstlisting}[style=JavaScript, language=C, xleftmargin=0.3cm, caption={Input-dependent code from $\tt{HardTanh.c}$ in Torch.
% from \texttt{THNN/generic/SpatialDilatedMaxPooling.c}.
},captionpos=b, label={lst:hardtanh}]
    real* ptr_input  = THTensor_(data)(input);
    real* ptr_output = THTensor_(data)(output);
    ptrdiff_t i;
    ptrdiff_t n = THTensor_(nElement)(input);

    if (inplace)
#pragma omp parallel for private(i)
      for (i = 0; i < n; i++){
        if (ptr_input[i] < min_val)
          ptr_input[i] = min_val;
        else if (ptr_input[i] > max_val)
          ptr_input[i] = max_val;
      }
    else
#pragma omp parallel for private(i)
      for (i = 0; i < n; i++){
        if (ptr_input[i] < min_val)
          ptr_output[i] = min_val;
        else if (ptr_input[i] <= max_val)
          ptr_output[i] = ptr_input[i];
        else
          ptr_output[i] = max_val;
\end{lstlisting}

%% file: listings/example.tex
\begin{lstlisting}[style=JavaScript, language=C, xleftmargin=0.3cm, caption={Example of an input-dependent branch.
% from \texttt{THNN/generic/SpatialDilatedMaxPooling.c}.
},captionpos=b, label={lst:example}]
void THNN_SpatialDilatedMaxPooling_updateOutput 
(...)  {
  ...
  if (val > maxval)
      maxval = val;
  ...
  /* set output to local max */
    *op = maxval;
\end{lstlisting}

%% file: chapters/design.tex
\section{\codename Design}
\label{sec:design}

In this section, we present an overview and the design details of
\codename which consists of \codename-Generator and
\codename-Converter that are trusted.\footnote{The code is small and
can be verified by inspection.}

\input{figures/overview.tex}

\subsection{Overview}
The model owner generates an inference model for the \codename system
using two steps (Figure~\ref{fig:overview}). In Step 1, the model
owner uses \codename-Converter to transform a deep learning framework
to be input-oblivious. In our prototype, we have designed
\codename-Converter to transform the popular Torch library. This is a
one-time process for any given framework.  \codename-Converter
identifies all the \pattern{} patterns in the DNN framework and
replaces them with the input-oblivious primitives. The converter is
generic with regards to the underlying oblivious primitives. Any 
oblivious construction can be plugged into the \codename-converter as
long as the construction guarantees obliviousness for all the branch
statements involving sensitive inputs. We select the $\tt{CMOV}$-based
solution that has been used to ensure obliviousness in previous
work~\cite{zerotrace,usenix16,raccoon}. 

In Step 2, the model owner uses \codename-Generator to generate the
model binary. \codename-Generator takes as input the ONNX
representation of the model, the input-oblivious DNN framework that
Step 1 generated, required SGX libraries, and an encryption key. It
outputs an enclave-executable model binary and the encrypted model
parameters. ONNX is an open neural network exchange format~\cite{onnx}
that is supported by several existing deep learning frameworks (e.g.,
Caffe2, Tensorflow, CNTK, and others).  Finally, the model owner
uploads the model binary to the cloud provider to host it as an
inference service, as described in Section~\ref{sec:setting}.

\input{chapters/converter}

\subsection{\codename-Generator} 
\label{subsec:o2e}

The \codename-Generator takes in an ONNX representation of a trained
model and outputs an enclave-executable model binary and encrypted
parameters. The ONNX format file contains the model
configuration details such as the layers, neurons as well as the
values for weights and biases.

\paragraph{Generating Enclave-specific Code.}
For a given model, \codename-Generator generates two pieces of code: 
one executes inside the enclave and the other executes outside.  It also generates an $\tt{edl}$ file which defines ecalls
(entry calls to the enclave) and ocalls (exit calls from the enclave).
Specifically, \codename-Generator defines two ecalls: 
$\tt{initialize()}$ initializes the parameters of the model and
$\tt{infer(image)}$ performs inference on encrypted images. It
defines one ocall, $\tt{predict()}$, which returns the model's
(encrypted) predicted class.  \codename-Generator generates
enclave-bound functions by parsing the ONNX model and converting the
ONNX operators to corresponding Torch function calls. For example,
some ONNX operators have an equivalent function in Torch (ReLU's
equivalent is $\tt{THNN\_FloatThreshold\_updateOutput()}$) while
other operations, like grouped convolution, require composing multiple
Torch functions in a for-loop. The non-enclave code deals with
creating and initializing the enclave, waiting on a socket to receive
client connections and  encrypted images, calling the ecall to perform
inference, and returning the predicted class to the client.

\paragraph{Reducing TCB.} 
\codename-Generator trims the Torch library to only the bare minimum
set of files required to compile a given model.  Once it identifies
all layers in the ONNX model, it includes and compiles only the
necessary Torch files from the math and the NN libraries. This step
excludes irrelevant library code and thereby reduces the TCB.

\paragraph{Encrypting Parameters.}
The model weights and parameters are encrypted by the model owner. The
owner shares the encryption key over a secure channel to the enclave
that executes the model binary as described in
Section~\ref{sec:setting}.

%

%% file: figures/overview.tex
\begin{figure}[t]
\centering
\includegraphics[scale=0.7]{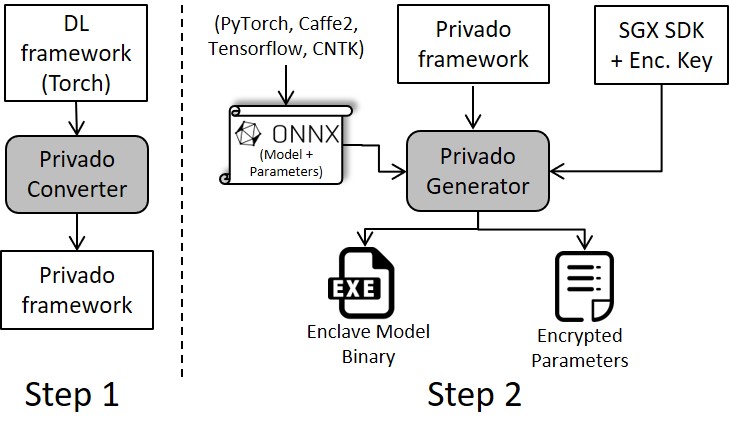}
\caption{
Overview of how the model owner generates the encrypted model and
parameters using the \codename-Converter and the \codename-Generator.
}
\label{fig:overview}
\end{figure}

%% file: chapters/converter.tex
\subsection{\codename-Converter}
\label{subsec:access-pattern}

In this section, we outline how \codename-Converter identifies
input-dependent conditions and makes them oblivious.

\paragraph{Identifying Input-dependent Branches.}
DNN frameworks consist of libraries that the model uses to construct
the final model binary.  \codename-Converter's first step is to
statically analyze the source code of these libraries and identify the
branches.  To do this, \codename statically traverses the AST of each
function in the library and reports conditional statements such as
if-else, input-dependent loop guards, and ternary operations.
\codename then performs an inter-procedural data flow analysis to
identify all the input-dependent variables~\cite{data-flow-analysis}.
Finally, \codename-Converter collects all the variables that are
involved in each conditional  statement and selects only the ones that
are input-dependent.

Next, \codename-Converter categorizes the input-dependent conditional
statements into {\em public} and {\em private} input-dependent
branches. Specifically, we white-list all the branches that use purely
public input, i.e., hyper-parameters such as network size, input size,
etc. Then, \codename-Converter performs taint 
analysis~\cite{confllvm}  with sensitive variables such as  query
input, weights, and biases marked as taint sources. \codename's
analysis propagates the taint and identifies the program variables in
the sinks i.e., the conditional statements of the deep learning
framework that leak private information. Thus \codename can check if
any of the branches use private input-dependent variables. It marks
all such branches that can potentially leak the user's private data.
Listing~\ref{lst:example} shows one such branch in the Torch library
implementation of the Max-pool layer, where the variables $\tt{val}$
and $\tt{maxval}$ used in the if-else condition on Line $4$ are
derived from the input. $\tt{val}$ and $\tt{maxval}$ are
activation values of neurons that are provided as input to the pooling
layer.

Our automated analysis provides us with a candidate list of private 
input-dependent branches, i.e., branches that may potentially leak 
inputs.  We study a sampled set of the branches flagged by our
analysis and observe that all of  them adhere to the \pattern{}
pattern. We do a best-effort source code analysis using a compiler
pass to identify well-known branching constructs such as loops,
if-else, and ternary operators. It is possible that our static
analysis may miss some branches (e.g., inline assembly code).   To
this end, we further empirically cross-checked our branch analysis by
comparing dynamic execution traces of the library for multiple inputs
in our evaluation. Our results confirm that our best-effort analysis
at least detects all the branches involved in the network
implementations that we test \codename on (See
Section~\ref{subsec:trace-equiv}). 

In the future, dynamic monitoring can ensure that \codename never
misses any branches that leak information,  because it did not
statically detect a branch in the library. Specifically, similar to
binary-instrumentation techniques for CFI, an enclave-level mechanism
can flag new branches whenever it observes that the input-dependent
execution traces are about to deviate the execution
trace~\cite{confllvm}. Thus, our present static analysis can be
combined with dynamic monitoring to ensure the completeness of our branch
analysis.

\paragraph{Using $\tt{CMOV}$ for Obliviousness.}
We use $\tt{CMOV}$ in a way similar to previous work~\cite{raccoon}.
$\tt{CMOV}$ is an x86 and x86-64 instruction that accepts a
condition, a source operand, and a destination operand. If the
condition is satisfied, it moves the source operand to the
destination. The important thing to note is that the $\tt{CMOV}$
instruction is oblivious when both the source and the destination
operands are in registers: it does not cause any memory accesses,
irrespective of the condition.

The key difference from previous work is that \codename-Converter
automatically modifies all occurrences of such branches in the DNN
library with equivalent code that uses $\tt{CMOV}$ with registers.
Moreover, using $\tt{CMOV}$ is just an implementation choice.
Alternatively, we can use other oblivious primitives such as
$\tt{AND}$ and $\tt{XOR}$ in \codename-Converter.
Let us consider the (simplified) branch code in
Listing~\ref{lst:patterns} that follows the \pattern{} pattern.
\input{listings/patterns}
Listing~\ref{lst:patterns-cmov} shows how \codename-Converter replaces
Listing~\ref{lst:patterns} with a functionally equivalent oblivious
code. 
\input{listings/patterns-cmov}
First, the code copies sensitive values $\tt{x}$ and $\tt{y}$ into
registers $\tt{eax}$ and $\tt{ebx}$ respectively. Then, the $\tt{
CMOV}$ instruction does a register-to-register copy based on the output
of the comparison instruction. Note that the adversary cannot observe
the register values in the secure CPU package. Therefore, using
$\tt{CMOV}$ instruction makes memory access patterns deterministic
and oblivious.  This ensures that after our transformation, the
library  does not leak any information. \codename-Converter employs
different conditional move instructions (e.g., $\tt{CMOVL}$,
$\tt{CMOVZ}$, $\tt{CMOVE}$, $\tt{CMOVBE}$, $\tt{FCMOVBE}$)
based on the specific instances of the branch patterns as well as the
data types of the variables. Listing~\ref{lst:example-fix} shows the
use of $\tt{FCMOVBE}$ instructions to replace the leaky branch in the
max-pool example in Listing~\ref{lst:example}.
\input{listings/example-fix}

\paragraph{Automated Transformation.}
\codename-Converter uses the LLVM compiler for the source-to-source
transformation of the input-dependent branches.  Thus, we automate the
transformation and ensure that the code uses $\tt{CMOV}$. Our analysis
ensures that the transformed code preserves the intended functionality
of the algorithm.  Since the number of branches which are transformed
is relatively small, we manually check that the transformed code is
functionally equivalent to the original code. We also empirically
verified this claim  re-running the inference to  ensure  that the
accuracies of the  models are indeed preserved after
\codename-Converter's transformation.


%
%


%% file: listings/patterns.tex
\begin{lstlisting}[style=JavaScript, language=C, xleftmargin=0.3cm, caption={Simple example of \pattern~ pattern.
%from \texttt{THNN/generic/SpatialDilatedMaxPooling.c}
},captionpos=b, label={lst:patterns}]
if (x < y)
	x = y;
\end{lstlisting}

%% file: listings/patterns-cmov.tex
\begin{lstlisting}[style=JavaScript, language=C, xleftmargin=0.3cm, caption={Oblivious code by replacing $\tt{<}$ operator with $\tt{cmovl}$ 
%from \texttt{THNN/generic/SpatialDilatedMaxPooling.c}
},captionpos=b, label={lst:patterns-cmov}]
//if (x < y) 	x = y;
mov eax, x
mov ebx, y
cmp eax, ebx  
cmovl eax, ebx  
mov x, eax       
\end{lstlisting}

%% file: listings/example-fix.tex
\begin{lstlisting}[style=JavaScript, language=C, xleftmargin=0.3cm, caption={Using conditional move instructions to hide input-dependent branch in Listing~\ref{lst:example}.
%from \texttt{THNN/generic/SpatialDilatedMaxPooling.c}
},captionpos=b, label={lst:example-fix}]
  float temp;
  asm volatile("fld %2 \n"
               "fld %3 \n"
               "fcomi \n"
               "fcmovbe %%st(1), %%st \n"
               "fstp %0 \n"
               "fstp %1 \n"
               :"=m"(maxval), "=m"(temp)
               :"m"(val), "m"(maxval));
\end{lstlisting}

%% file: chapters/evaluation.tex
\section{Evaluation}
\label{eval}

In this section, we first provide a brief outline of the \codename
implementation. Next, we state our evaluation goals. Finally, we
describe the results of our evaluation.

\input{tables/benchmarks_tbl.tex}

\subsection{Implementation}

Our \codename prototype is built on the Torch DNN framework. 
\codename-Converter is implemented as an LLVM source-to-source
transformation pass and has $1437$ lines of C code. 
\codename-Generator  has $1700$ lines of C++ code and uses the 
$\tt{sgx}$-$\tt{crypto}$ library to implement the
encryption-decryption functions in the enclave. We use Intel SGX
SDKv.2 for Linux and compile using GCCv5.4 with -$\tt{02}$
optimization flag. We select $11$ state-of-the-art models for
our experiments (Table~\ref{tbl:benchmarks}).

\input{tables/performance_tbl}

\subsection{Evaluation Goals}

Our evaluation answers the following questions:
\begin{itemize}
\squish
\item  {\em Ease-of-use.} 
How easy is it to use \codename to generate secure versions of
various state-of-the-art models?

\item {\em Performance.} 
How much overhead does \codename add, compared to baseline (insecure)
inference?

\item  {\em Lowering TCB.} 
How much TCB does \codename reduce?

\item {\em Obliviousness.} 
Are the generated models oblivious?
\end{itemize}

\subsection{Ease-of-use}
\label{sec:ease}
In our experiments, we used \codename-Generator on $11$
state-of-the-art neural network models specified in ONNX. We select
these models based on the differences in their depth (or the number of
layers), parameter sizes, training dataset, and accuracy as shown in
Table~\ref{tbl:benchmarks}.  \codename-Generator successfully
transformed all these trained models from their given ONNX format to
SGX-enabled code within a few seconds. No custom-coding was required.
We chose networks that achieve good accuracy on three popular datasets
for images: MNIST~\cite{mnist}, Cifar-10~\cite{cifar}, and
ImageNet~\cite{imagenet}.  LeNet uses the least number of parameters
($62$K), and AlexNet uses the highest number ($61.1$M).  The number of
layers in these models range from $6$ (MLP), all the way to $910$
(DenseNet). Lastly, \codename does not reduce the prediction accuracy
because (a) the models are trained to convergence in a trusted
environment before they are uploaded to the cloud; (b) the \codename
transformation does not change the model inference technique or the
specific weights and biases. We experimentally confirm that
\codename-Generator and \codename-Converter preserve the accuracy of
the model; they yields similar accuracies as the
original ONNX models on the target datasets.


\subsection{Performance}
\label{sec:performance}

\paragraph{Experimental Setup.}
We run all our experiments on a machine with  6th Generation Intel(R)
Core(TM) i7-6700U processor, $64$GB RAM, and $8192$ KB L3 cache with
Ubuntu Desktop-16.04.3-LTS $64$ bits and Linux kernel version
4.15.0-33-generic as the underlying operating system. The BIOS is
configured to use the $128$ MB for SGX. All our measurements are
averaged over $100$ iterations and use a single core.

\paragraph{Methodology.} 
For every model, we compute three execution time metrics. The {\em
baseline} time measures standard versions of the model running outside
SGX. The {\em SGX} time measures time to run the model within SGX, but
with no obliviousness.  The {\em SGX+CMOV} time measures time to run
the model within SGX with added support for obliviousness.  During
each execution, we record statistics such as the enclave memory size,
number of page faults, and input-dependent branches for each of the
models. Table~\ref{tbl:performance} shows the execution time for our
benchmark models. \codename incurs overhead between $-27.4\%$ and
$90.4\%$. We believe these numbers are acceptable,  given the
additional security and privacy guarantees. We now describe several
interesting results from our evaluation.

\paragraph{SGX-Enclaves improve efficiency for models that fit
entirely in SGX memory.} 
Surprisingly, our first finding is that {\it some models, namely
MLP, LeNet, VGG19, Resnet110, and Squeezenet, execute faster with
\codename than the baseline }(negative overhead in
Table~\ref{tbl:performance}). Note that these models are relatively
small: they fit entirely within SGX memory (90~MB), and so do not
incur page fault overheads. Upon further analysis, we observed that
these models run faster with SGX because the SGX SDK provides
an efficient implementation of some $\tt{libc}$ functions, such as
$\tt{malloc}$, as compared to standard $\tt{libc}$ used in the
baseline execution.

\paragraph{Page faults cause most of the overhead.}  
For models that exceed the 90 MB SGX memory limit, we find that
page swapping between the enclave and outside memory (page faults)
contributes to most of the performance overhead. To explain this, we
use the metric {\em normalized page faults per second} (NPPS) in
Table~\ref{tbl:performance}. This metric calculates the number of
page faults incurred by the enclave while executing model inference
for one image, normalized by the baseline execution time. We observed
that the performance overhead is highly correlated with the NPPS
metric for all models. Among models that exceed $90$MB, AlexNet has
the highest NPPS ($65245$) and the highest overhead ($90.4$\%) while
Wideresnet has the lowest NPPS ($5080$) and lowest overhead
($15.6$\%). These results suggest that for SGX inference, there are
{\it optimization opportunities in model design and implementation
that can carefully minimize memory size and/or page faults}.

\paragraph{Obliviousness is cheap for neural networks.} 
Observe columns {\em SGX} and {\em SGX+CMOV} in
Table~\ref{tbl:performance}. The former captures SGX overheads without
obliviousness (including decryption of input which costs only about
$84$ and $313$ microseconds for Cifar-10 and ImageNet images
respectively), while the latter captures SGX overheads after adding
obliviousness.  We find that {\it the overhead of adding obliviousness
is a mere $1\%$ on average for these ten DNN models}.  This
observation confirms our claim that neural networks have relatively
few input-dependent branches and that the bulk of the computation is
access independent, and hence the cost of achieving obliviousness is
negligible.

\input{figures/batchsize}

\paragraph{Batching may be counterproductive.}
We perform an additional experiment to understand the performance
overhead for inferring multiple images in a batch. Batching is
typically used to improve inference performance.
Figure~\ref{fig:batchsize} shows the execution time for inferring
images in batch sizes of one to four for the Squeezenet model. As
shown in Table~\ref{tbl:performance}, Squeezenet incurs a negative
overhead for a single image as compared to the baseline execution.
However, as the batchsize increases, we observe that the execution
time for \codename exceeds that of baseline for batchsize of three
images and onwards. An increase in the batch size results in larger
enclave memory size as the number of activations increases
proportionately with batch size. For a batch size of three, Squeezenet
memory usage exceeds 90~MB and thus the overhead switches from
negative to positive. This result suggests the following rule of
thumb: {\it while performing inference in SGX enclaves, to minimize
performance overhead, a smaller batch size that limits memory usage
below $90$MB is essential}.

\input{figures/trace_comparison}

\paragraph{Comparison to Previous Work.}
We compare \codename to previous work by \olya~\cite{usenix16}
in terms of performance overhead, expressiveness, and mechanism
for obliviousness. We observe that the overhead of \olya due to SGX
and encryption ranges from $0.01$\% to $91$\% while that for \codename
ranges from $-27.4$\% to $90.4$\%. Similarly, for the neural network
model on MNIST dataset, the overhead is around $ 0.3$\% due to
obliviousness for them which is the same (i.e., $0.3519$\% with an 
absolute execution time of $0.000571$ sec) for the MLP model with
\codename (as shown in Table~\ref{tbl:performance}).   The overhead
due to obliviousness for each model depends on the number of
data-dependent layers and the neurons in each layer present in that
model. Next, \olya  demonstrate expressiveness by manually modifying
$5$ distinct machine learning algorithms selected from different
categories  while \codename focuses specifically on deep learning
algorithms and is more expressive for this class of neural
networks.  \codename automatically converts any given convolutional
neural network to run obliviously in the enclave as we modify all the
data-dependent layers including pooling layers that are not handled in
previous work. At last, although the main technique in \codename to
ensure obliviousness is similar in several previous
works~\cite{raccoon, usenix16, zerotrace, obliviate}, we leverage our
key observation of {\em assign-or-nothing} patterns in neural networks
to design a fully-automated system. Thus, our novel method of 
applying the oblivious primitive to neural networks combined with
\codename-Converter and \codename-Generator is the key contribution of
this paper. Moreover, the techniques in \codename are not limited to
convolutional neural networks but are suitable for neural networks
such as RNNs or LSTMs that exhibit similar data-dependent access
patterns.

\subsection{Lowering TCB}
\label{sec:eval-tcb}

A naive implementation of \codename would require trusting the entire
Torch math and NN libraries, which have approximately $30,000$ lines
of code. However, \codename-Generator further lowers the TCB using the
technique described in Section~\ref{subsec:o2e}.  To calculate the
reduction in the TCB, we count the number of lines used to generate
each of the $10$ model binaries.  The last column, ``TCB Reduction'',
in Table~\ref{tbl:performance} shows the percentage reduction in TCB
as compared to trusting the entire torch library of $30,000$ lines for
each model.  On average, \codename~ {\it results in a $34.7$\%
reduction in TCB for our benchmark models}.

\subsection{Obliviousness}
\label{subsec:trace-equiv}

\codename-Converter transforms the program to ensure that the
execution trace is the same for all inputs. To empirically evaluate
this claim, we trace all the instructions executed by the model binary
and record all the memory read/write addresses.  We build a tracer
using a custom PinTool based on a dynamic binary instrumentation tool
called Pin~\cite{pin}. Our tracer logs each instruction before it is
executed, along with the  memory address accessed (read/write) by the
instruction.  We run our models before and after using
\codename-Converter and log the execution traces for all the inputs in
our dataset for checking obliviousness.

Figure~\ref{fig:trace_comparison}~(a) shows two execution trace
snippets of LeNet on two different inputs with labels ($\tt{airplane}$
and $\tt{horse}$) before applying \codename-Converter} (same as
Section~\ref{sec:attack_details}, 
Figure~\ref{fig:attack_feature_trace}). Observe that the left-hand
side trace executes more and different instructions that access
different addresses in different modes (read / write) as compared to
the right-hand side trace. The adversary can thus distinguish between
two inputs by observing such differences in the execution trace.  This
empirically reinstates that existing library implementations indeed
leak input information.

Figure~\ref{fig:trace_comparison}~(b) shows the traces after
\codename-Converter, they are identical after the  transformation. 
This confirms that \codename-Converter fixed the branches which leaked
information in Figure~\ref{fig:trace_comparison}~(a). As an empirical
evaluation, we collect such execution traces for all the models and
their corresponding inputs in the dataset and check the traces. We
report that in our experiments, we do not detect any deviation. Hence,
 we confirm experimentally that we did not miss any branches at least
for all the models  and inputs in our evaluation.

%% file: tables/benchmarks_tbl.tex
\begin{table}[t]
  \centering
  \small
\begin{tabular}{|l|l|l|c|l|}
\hline
\textbf{Models} & \textbf{\begin{tabular}[c]{@{}c@{}}No. of \\ Layers\end{tabular}} & {\textbf{\begin{tabular}[c]{@{}c@{}}No. of \\ Param.\end{tabular}}} & {\textbf{Dataset}} & {\textbf{\begin{tabular}[c]{@{}c@{}} Acc. (\%)\\Top 1/ 5\end{tabular}}} \\  \hline
MLP*       & 6                      & 538K                          & MNIST                  & 99                       \\ \hline
LeNet*          & 12                      & 62 K                          & Cifar-10                 & 79                      \\ \hline
VGG19          & 55                      & 20.0 M                          & Cifar-10                 &92.6                        \\ \hline
Wideresnet*       & 93                      & 36.5 M                          & Cifar-10                 & 95.8                       \\ \hline
Resnext29*      & 102                      & 34.5 M                          & Cifar-10                 & 94.7                       \\ \hline
Resnet110*      & 552                      & 1.70 M                          & Cifar-10                 & 93.5                       \\ \hline \hline
AlexNet       & 19                      & 61.1 M                          & Imagenet                 & \hspace{1.0cm}80.3                       \\ \hline
Squeezenet     & 65                      & 1.20 M                          & Imagenet                 & \hspace{1.0cm}80.3                       \\ \hline
Resnet50       & 176                      & 25.6 M                          & Imagenet                 & \hspace{1.0cm}93.6                      \\ \hline
Inceptionv3      & 313                      & 27.2 M                         & Imagenet                  & \hspace{1.0cm}93.9                       \\ \hline
Densenet       & 910                      & 8.10 M                          & Imagenet                  & \hspace{1.0cm}93.7                       \\ \hline

\end{tabular}
\caption{ Models evaluated with \codename. Columns $1-5$ show the model name, the total number of layers, the number of parameters,
the dataset and the accuracy respectively. * indicates we trained
these models ourselves; rest of the models were  taken from ONNX
repository of pre-trained models~\cite{onnx-model-source}. We show Top
1 and Top 5 accuracies we observed in our inference for MNIST,
Cifar-10, and ImageNet datasets.}
\label{tbl:benchmarks}
\end{table}

%% file: tables/performance_tbl.tex
\begin{table*}[t]
  \centering
\begin{tabular}{|c|r|r|r|r|r|r|c|}
\hline
\multirow{2}{*}{\textbf{Models}} & \multicolumn{3}{c|}{\textbf{Execution time (in sec)}}                                                  & \multirow{2}{*}{\textbf{\begin{tabular}[c]{@{}c@{}}Overhead\\(in \%)\end{tabular}}} & \multirow{2}{*}{\textbf{\begin{tabular}[c]{@{}c@{}}Enclave\\ mem size \\ (in MB)\end{tabular}}} & \multirow{2}{*}{\textbf{\begin{tabular}[c]{@{}c@{}}\ Normalized \\ Page-Fault \\ per sec \end{tabular}}}  & \multirow{2}{*}{\textbf{\begin{tabular}[c]{@{}c@{}}TCB\\ Reduction \\ (in \%)\end{tabular}}} \\ \cline{2-4}
                                 & \textbf{Baseline}      & \textbf{{\begin{tabular}[c]{@{}c@{}}SGX \end{tabular}}}     &
					  \textbf{{\begin{tabular}[c]{@{}c@{}}SGX + \\ CMOV\end{tabular}}}     &                                                                                      &                                                                                                &                                                                                                                                                                                                      &                                                                                              \\ \hline
\begin{tabular}[c]{@{}c@{}}MLP\end{tabular}                    & 0.0007 & 0.000569  &  0.000571 &           -18.7                                                      &  3.08                                                                  & 0                                                                                                                                                      &   44.9                                                                   \\ \hline
\begin{tabular}[c]{@{}c@{}}LeNet\end{tabular}                           & .001                    & .0007                                           & 0.000726                      & -27.4                                                                                     & 0.624                                                                                        & 0                                                                                                                                                                                                 & 34.4                                                                                             \\ \hline
\begin{tabular}[c]{@{}c@{}}VGG19\end{tabular}                           & 0.62                      & 0.61                                           & 0.61136                      & -1.4                                                                                     & 89.7                                                                                       & 0                                                                                                                                                                                                  & 33.2                                                                                             \\ \hline
\begin{tabular}[c]{@{}c@{}}Wideresnet\end{tabular}                      & 12.3                      & 14.2                                           & 14.2252                      & 15.6                                                                                     & 248.8                                                                                       & 5080                                                                                                                                                                                                    & 33.4                                                                                             \\ \hline
\begin{tabular}[c]{@{}c@{}}Resnet110 \end{tabular}                       & 0.38                      & 0.32                                         & 0.3242                      & -14.6                                                                                     & 81.3                                                                                       & 0                                                                                                                                                                                                   & 33.4                                                                                             \\ \hline
\begin{tabular}[c]{@{}c@{}}Resnext29\end{tabular}                        & 9.14                      & 10.9                                           & 10.907                      & 19.3                                                                                     & 267.9                                                                                       & 7302                                                                                                                                                                                                 & 33.4                                                                                             \\ \hline
\begin{tabular}[c]{@{}c@{}}AlexNet\end{tabular}                          & 0.99                      & 1.88                                           & 1.885                      & 90.4                                                                                     & 258.0                                                                                       & 65245                                                                                                                                                                                                 & 33.8                                                                                             \\ \hline
\begin{tabular}[c]{@{}c@{}}Squeezenet\end{tabular}                        & 0.49 & 0.38 & 0.387 & -26.6                                                                & 48.0                                                                  & 0                                                                                                                                                         & 34.2                                                                        \\ \hline
\begin{tabular}[c]{@{}c@{}}Resnet50\end{tabular}                         & 4.51                      & 6.19                                            & 6.207                     & 37.6                                                                                    & 332.1                                                                                       & 18412                                                                                                                                                                                                   & 33.7                                                                                             \\ \hline
\begin{tabular}[c]{@{}c@{}}Inceptionv3\end{tabular}                     & 6.56 & 8.64 & 8.646 & 31.7                                                                & 391.1                                                                  & 14935                                                                                                                                                        & 33.9                                                                         \\ \hline
\begin{tabular}[c]{@{}c@{}}Densenet\end{tabular}                    & 3.08 & 5.60  &  5.639 &           83.1                                                      &  547                                                                  & 44439                                                                                                                                                      &   33.5                                                                   \\ \hline

\end{tabular}
\caption{ Performance for inferring a single image with \codename on ten DNN models on a single-core: LeNet~\cite{lenet}, VGG19~\cite{vgg}, Wideresnet~\cite{wideresnet}, Resnet110~\cite{resnet}, Resnext29~\cite{resnext}, AlexNet~\cite{alexnet}, Squeezenet~\cite{squeezenet}, Resnet50~\cite{resnet}, Inceptionv3~\cite{inception}, and Densenet~\cite{densenet}.
Columns $2-5$ show the performance breakdown, negative value implies that \codename execution time is less than the baseline.  Column $6$ shows the size of the maximum memory footprint of the enclave observed at runtime.
Column $7$ shows the NPPS and Column $8$ represents the reduction in the TCB  w.r.t. to the original Torch library code.
}
\label{tbl:performance}
\end{table*}

%% file: figures/batchsize.tex
\begin{figure}[t]
	\centering
	\includegraphics[scale=0.42]{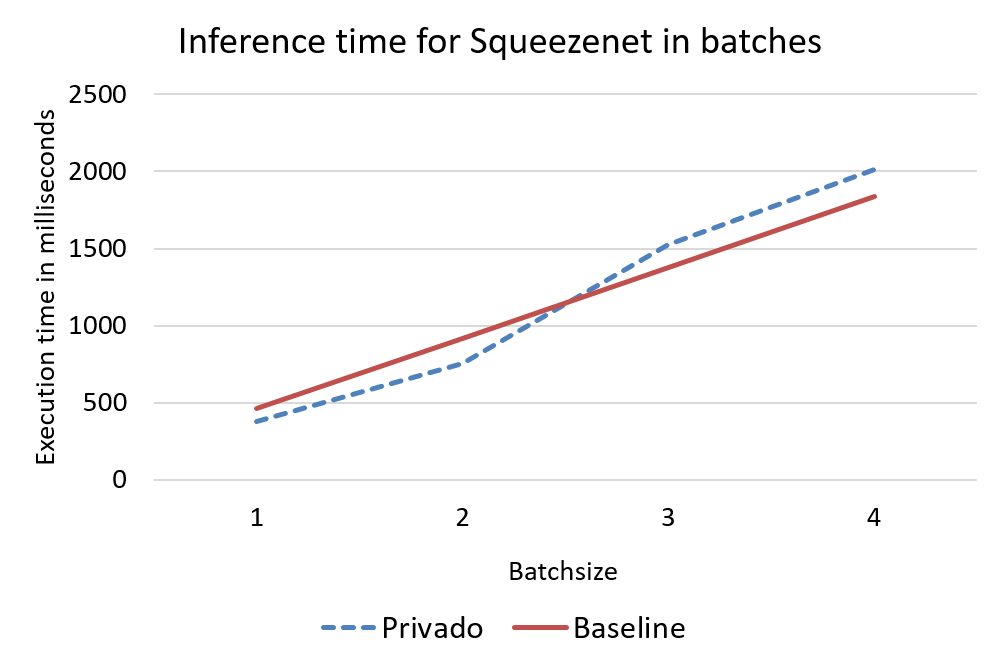}
	\caption{ Execution time vs. batch size plot shows that Squeezenet 
	execution time overhead increases with increase in the batch size.}
	\label{fig:batchsize}
\end{figure}

%% file: figures/trace_comparison.tex
\begin{figure*}[t]
\centering
\includegraphics[scale=0.57]{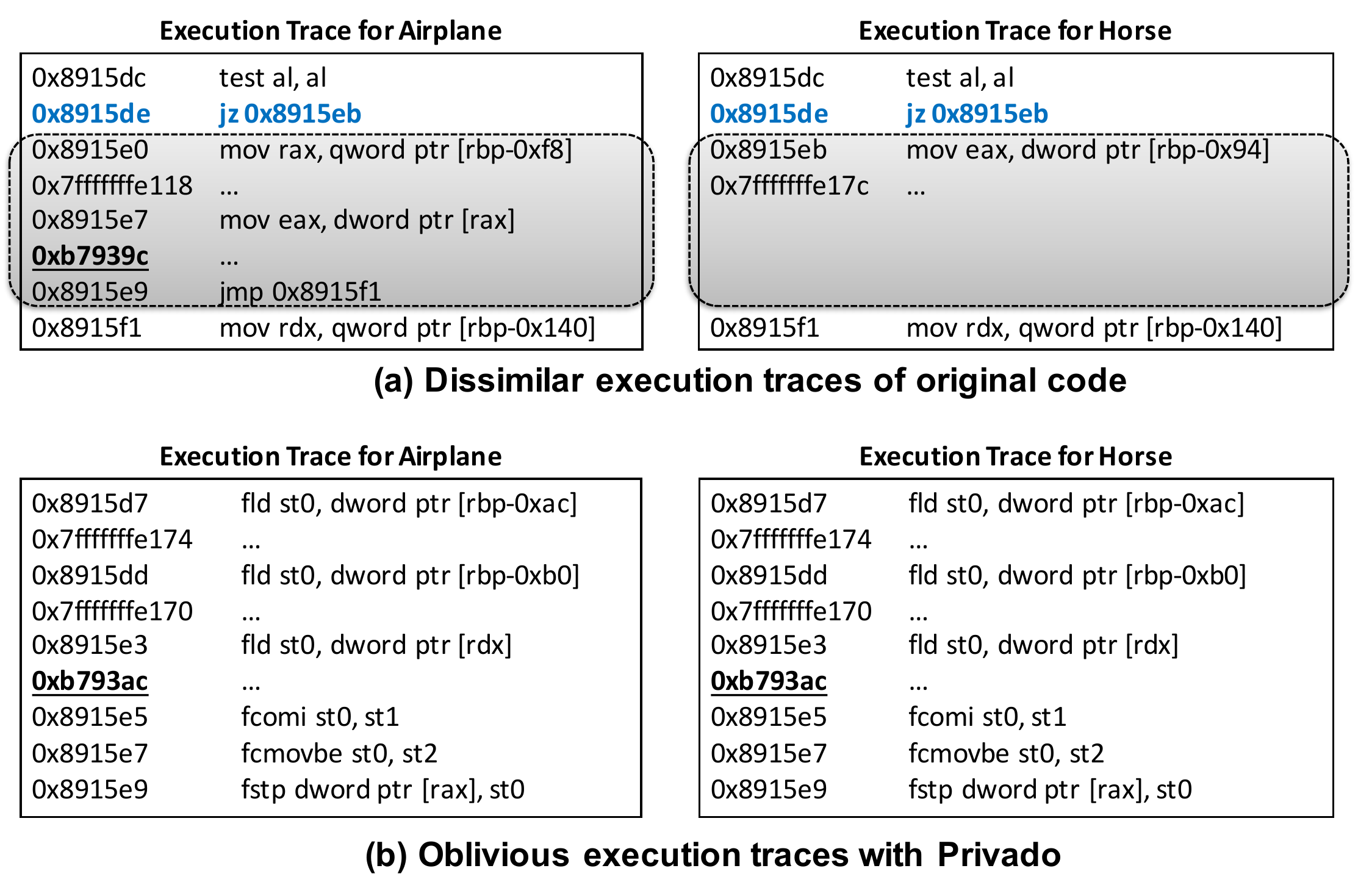}
\caption{
Execution traces for LeNet on two different inputs (a) before and (b)
after applying \codename. The underlined code in (a) shows a
conditional jump ($\tt{jz}$). The highlighted trace snippet depicts
the assign-or-nothing pattern executed for Airplane but not for Horse.
The bold and underlined part shows the address which causes a
page-fault that the OS can use to distinguish if the branch was
executed or not. (b) After \codename transformation, the traces are
 the same and hence oblivious.
}
\label{fig:trace_comparison}
\end{figure*}




%% file: chapters/related.tex
\section{Related Work}
\label{sec:related}

In this section, we discuss prior-work on secure neural-network
inference using: (a) cryptographic primitives such as homomorphic
encryption; and (b) trusted hardware.

\subsection{Cryptographic Primitives}
CryptoNets~\cite{cryptonets} was the first to use homomorphic
encryption to support neural network inference on encrypted data. 
Following Cryptonets, several other solutions such as
DeepSecure~\cite{deepsecure}, Minionn~\cite{minionn},
SecureML~\cite{secureml}, ABY3~\cite{mohasselaby3},
SecureNN~\cite{waghsecurenn}, and Gazelle~\cite{gazelle}, 
Chameleon~\cite{riazi2018chameleon}, Tapas ~\cite{sanyal2018tapas} 
have been proposed for secure neural network inference. These
solutions use a combination of cryptographic primitives such as
garbled circuits, secret-sharing, and fully homomorphic encryption.
Thus, the solutions provide stronger security guarantees as they do not
trust any hardware entity. However, these solutions use heavy-weight
cryptography and hence  incur a significant performance overhead while
limiting the ease of use thus making them difficult to adopt in
practice.

\codename takes an orthogonal approach to these solutions and uses
trusted hardware as the main underlying primitive.  Although trusted
hardware-based approaches are not a silver bullet and have their 
share of limitations, we used them because they are  more suitable for
practical deployments. In comparison to cryptographic primitive-based
solutions, \codename provides  practical execution performance relying
on the availability of trusted hardware and the correct
implementation of a secure channel and attestation mechanisms. The
choice of using cryptographic primitives or trusted hardware solutions
depend on the trade-off that is acceptable between security and
performance guarantees.

\subsection{Trusted Hardware}
\olya propose a customized oblivious solution for machine learning
algorithms using SGX~\cite{usenix16}. As discussed in
Section~\ref{sec:performance}, it does not address the ease-of-use
challenge.  Recent work Myelin proposes the use of SGX primarily to
secure the training process using differential privacy to achieve
obliviousness guarantees~\cite{myelin}. \codename, on the other hand,
focuses on supporting end-to-end inference-as-a-service for a given
trained model. Further, unlike \codename, Myelin is not
backward-compatible i.e., it cannot execute inference for models that
are {\em not} trained using the Myelin framework. Our use of the
popular Torch framework and ONNX along with automated tools for
ensuring obliviousness with ease-of-use differentiates \codename from
previous work.

Hunt \etal~ proposed Chiron, a privacy-preserving machine learning
service using the Theano library and SGX~\cite{chiron}. Chiron does
not prevent leakage via access patterns which is a serious concern as
shown in Section~\ref{sec:attack}. Similarly, MLCapsule, a system for
secure but offline deployment of ML as a service on the client-side is
susceptible to leakage of sensitive inputs via access
patterns~\cite{mlcapsule}. Lastly, Slalom combines SGX and GPU for
efficient execution of neural network inference but also does not
address access-pattern leakage~\cite{slalom}.

Finally, in the light of various side-channel attacks on Intel SGX,
several recent works have proposed point-wise solutions to each source
of leakage via software and / or hardware
modifications~\cite{fu2017s,chen2017detecting, shih2017t,
oleksenko2018varys, sanctum, cosmix}. The main goal of these defenses
is to be generic-enough to protect a large class of enclave
applications at the cost of performance degradation, increased
developer-effort, or limited expressiveness of enclave-bound
applications. \codename, on the other hand, is tailored  for a
specific class of enclave applications---deep neural net inference. 
Our scope and problem setting allow us to strike a balance between
various design trade-offs. Several other proposals to thwart
side-channel leakage  cover a larger attack surface and hence can
prevent part of the leakage that we demonstrate via our \pattern{}
pattern~\cite{werner2019scattercache}. However, \codename assumes a
worst-case adversary and shows that even then it can eliminate such
leakage by making the enclave access-pattern oblivious.

%% file: chapters/conclusion.tex
\section{Conclusion}
In this work, we demonstrate the first-of-its-kind concrete attack to
highlight the importance of hiding access patterns in DNN inference.
Using real-world models executing on a real system, we show that an
adversary that observes enclave access patterns can predict encrypted
inputs with $97$\% and $71$\% accuracy for MNIST and CIFAR10 datasets
respectively. To this end, we present \codename---a system which
provides secure DNN inference. \codename is input-oblivious, easy to
use, requires no developer effort, has low TCB, and has low
performance-overheads. We implement \codename on the Torch framework,
apply it to $11$ contemporary networks, and demonstrate that \codename
is practical and secure.